\documentclass[aps,pra,twocolumn,showpacs,superscriptaddress,10pt,longbibliography]{revtex4-2}

\usepackage[colorlinks = true,linkcolor = red,citecolor = magenta]{hyperref}
\usepackage[sort&compress]{natbib}
\usepackage{scalerel}
\usepackage[normalem]{ulem}
\usepackage{xcolor}
\usepackage{bm}
\usepackage{amsmath}
\usepackage{amsfonts}
\usepackage{amssymb}
\usepackage{graphicx}
\usepackage{subfigure}
\usepackage{mathtools}
\usepackage{physics}
\usepackage{dsfont}
\usepackage{float}
\usepackage{lmodern}
\usepackage[many]{tcolorbox}
\usepackage{empheq}
\usepackage{cleveref}
\usepackage{textcomp}

\begin{document}

\title{Signatures of Quantum Chaos and fermionization in the incoherent transport of bosonic carriers in the Bose-Hubbard chain}
\author{P. S. Muraev}
\affiliation{Kirensky Institute of Physics, Federal Research Centre KSC SB RAS, 660036, Krasnoyarsk, Russia}
\affiliation{School of Engineering Physics and Radio Electronics, Siberian Federal University, 660041, Krasnoyarsk, Russia}
\affiliation{IRC SQC, Siberian Federal University, 660041, Krasnoyarsk, Russia}
\author{D. N. Maksimov}
\affiliation{Kirensky Institute of Physics, Federal Research Centre KSC SB RAS, 660036, Krasnoyarsk, Russia}
\affiliation{IRC SQC, Siberian Federal University, 660041, Krasnoyarsk, Russia}
\author{A. R. Kolovsky}
\affiliation{Kirensky Institute of Physics, Federal Research Centre KSC SB RAS, 660036, Krasnoyarsk, Russia}
\affiliation{School of Engineering Physics and Radio Electronics, Siberian Federal University, 660041, Krasnoyarsk, Russia}

\date{\today}

\begin{abstract}
We analyse the stationary current of Bose particles across the Bose-Hubbard chain connected to a battery, focusing on the effect of inter-particle interactions. It is shown that the current magnitude drastically decreases as the strength of inter-particle interactions exceeds the critical value which marks the transition to quantum chaos in the Bose-Hubbard Hamiltonian. We found that this transition is well reflected in the non-equilibrium many-body density matrix of the system. Namely, the level-spacing distribution for eigenvalues of the density matrix changes from Poisson to Wigner-Dyson distributions. With the further increase of the interaction strength, the Wigner-Dyson spectrum statistics changes back to the Poisson statistics which now marks fermionization of the bosonic particles. With respect to the stationary current, this leads to the counter-intuitive dependence of the current magnitude on the particle number.    
\end{abstract}
\maketitle


{\em 1.} Recently, we have seen a surge of interest to quantum transport in one-dimensional systems coupled at their edges to particle reservoirs. Following~\cite{Land21} we refer to these systems as boundary driven systems. Particularly, one very important example of the boundary driven system is the so-called open Bose-Hubbard (BH) model or BH chain \cite{Ivan13,112,116,123}. This system can be realised experimentally by using different physical platforms including superconducting circuits \cite{Fitz17,Fedo21}, photonic crystals \cite{Szam05,Cace22}, and cold Bose atoms in optical lattices \cite{Labo16}. The central question to be addressed with the open BH chain, both theoretically and experimentally, is the stationary current of Bose particles across the chain and its dependence on the strength of inter-particle interactions. It is known that properties of the closed/conservative BH system crucially depend on the ratio of the hopping matrix element $J$ and the interaction constant $U$ which are two of the four parameters of the BH Hamiltonian, the other being the chain length $L$ and the particle number $N$. For example, for integer $N/L$ the ground state of the system shows the quantum phase transition between the super-fluid state for $J\gg U$ and the Mott-insulator state in the opposite limit \cite{Grei02}. As for the excited states they show a qualitative change from the regular to the chaotic \cite{66,103}. Thus, one may expect that the stationary current in the open BH chain may also crucially  depend on the interaction constant.

Up to now the above problem has been analysed only by using the pseudoclassical approach which approximates the quantum dynamics by that of the classical counterpart of the BH model \cite{116}. This is because the exact quantum treatment, both numerical and analytic, is complicated by the fact that the open BH model does not conserve the particle number. Thus, the system dynamics takes place in the extended Hilbert space spanned by the direct sum of the subspaces with given number of particles. Taking into account that the transition to quantum chaos in the closed BH model depends not only on the ratio $U/J$ but also on the ratio $N/L$, it is very problematic to track this transition to in the transport properties of the open BH model.

In the present work we introduce a boundary driven BH model which shows a behaviour similar to the standard open BH model but conserves the number of particles. This allows us to separate the dependencies on the interaction constant $U$ and the particle number $N$, thus, relating the obtained results to the transition to chaos in the conservative BH model.  


{\em 2.} We consider the BH chain of the length $L$ with incoherent coupling between the first and the $L_{{\rm th}}$ sites. The coupling is described by the the following Lindblad operators,
\begin{equation}
  \begin{array}{cc}
  {\cal L}_1(\widehat{R}) &= 
  \widehat{V}^{\dagger}\widehat{V}\widehat{R} + \widehat{R}\widehat{V}^{\dagger}\widehat{V} 
   - 2\widehat{V}\widehat{R}\widehat{V}^{\dagger}
   \;, \\
   \\
   {\cal L}_2(\widehat{R}) &= 
   \widehat{V}\widehat{V}^{\dagger}\widehat{R} + \widehat{R}\widehat{V}\hat{V}^{\dagger} 
   - 2\widehat{V}^{\dagger}\widehat{R}\hat{V}
   \;,
\end{array}
\end{equation}
where $\widehat{V}=\hat{a}_1^\dagger \hat{a}_L$. Thus, the master equation for the carriers density matrix $\widehat{R}$ reads
\begin{equation}
\label{a2}
\frac{\partial \widehat{R}}{\partial t}=-i[\widehat{H}, \widehat{R}] -\Gamma_1{\cal L}_1(\widehat{R}) - \Gamma_2 {\cal L}_2(\widehat{R}) \;,
\end{equation}
where 
\begin{equation}
\label{a1}
\widehat{H} = -\dfrac{J}{2}\sum^{L - 1}_{\ell = 1}\left(\hat{a}^{\dagger}_{\ell + 1}\hat{a}_{\ell} + \mathrm{h.c.}\right) 
+ \dfrac{U}{2}\sum^L_{\ell = 1} \hat{n}_{\ell}(\hat{n}_{\ell}-1) 
\end{equation}
is the Bose-Hubbard Hamiltonian. It is easy to see that the Lindblad operator ${\cal L}_1(\widehat{R})$ induces the incoherent transport of the carriers from the last to the first sites, while the operator ${\cal L}_2(\widehat{R})$ is responsible for the incoherent transport in the reverse direction. If the rates $\Gamma_1 \ne \Gamma_2$, there is a non-zero current in the clockwise or counterclockwise direction depending on the inequality relationship between the two relaxation constants. 
We notice that, by an analogy with electronic devices, the introduced Lindblad operators mimic the effect of a battery which induces direct current in the electric circuits.


{\em 3.} We are interested in the stationary current $I={\rm Tr}[\widehat{I}\widehat{R}]$ where $\widehat{R}=\widehat{R}(t\rightarrow\infty)$ is now the steady-state density matrix and $\widehat{I}$ is the current operator,
\begin{equation}
\label{b3}
\widehat{I}= \dfrac{J}{2i} \sum^{L - 1}_{\ell = 1} \left(\hat{a}^{\dagger}_{\ell + 1}\hat{a}_{\ell} - h.c.\right) \;.
\end{equation}
First of all, we notice that if $\Gamma_1 = \Gamma_2$ the steady-state density matrix is proportional to the identity matrix, namely, $\widehat{R}=\widehat{1}/{\cal N}$, where ${\cal N}$ is the dimension of the Hilbert space. In what follows we focus on the liner response regime where $\Gamma_1=\Gamma +\Delta\Gamma/2$, $\Gamma_2=\Gamma -\Delta\Gamma/2$, and $\Delta\Gamma \ll \Gamma$. Thus, we have 
\begin{equation}
\label{b1}
\widehat{R}=\frac{\widehat{1}}{{\cal N} }+\Delta\Gamma \widetilde{R} \;,
\end{equation}
where ${\rm Tr}[\widetilde{R}]=0$. Substituting the Ansatz~(\ref{b1}) into the master equation we obtain 
\begin{equation}
\label{b2}
-i[\widehat{H},\widetilde{R}] - \Gamma \left[{\cal L}_1(\widetilde{R}) + {\cal L}_2(\widetilde{R})\right]
-\frac{2(\hat{n}_L-\hat{n}_1)}{{\cal N}} = O(\Delta\Gamma) \;.
\end{equation}
In the limit $\Delta\Gamma \rightarrow 0$ Eq.~(\ref{b2}) transforms into the algebraic equation for the elements of the unknown matrix $\widetilde{R}$. In our numerical approach, however, we do not solve this algebraic equation but evolve the density matrix $\widehat{R}(t)$ according to the master equation~(\ref{a2}) 
and use Eq.~(\ref{b2}) to check that we reached the true steady state. We found this method to be more efficient than the straightforward solution of the algebraic equation. 
\begin{figure}
\includegraphics[width=8.0cm,clip]{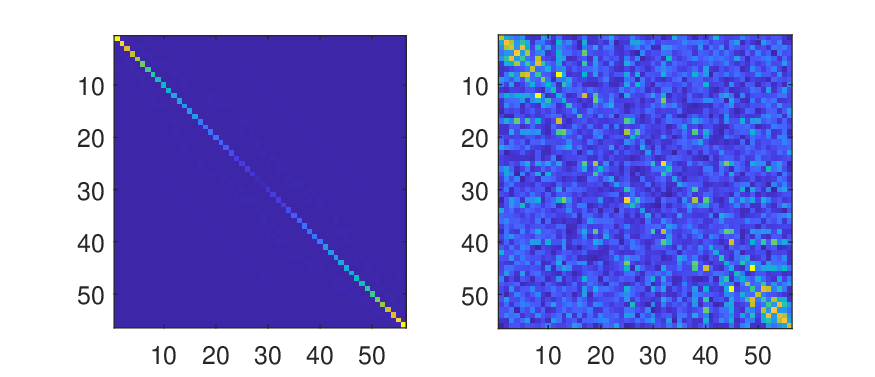}
\caption{Absolute values of the matrix elements of the matrix $\widetilde{R}$ in the basis of the current operator. The system parameters are $L=6$, $N=3$ (the Hilbert space dimension  ${\cal N}=56$), $J=1$, $\Gamma=0.04$, and $U=0$ (left) and $U=1$ (right). The upper limit of the colour axis is $0.2$.}
\label{fig3}
\end{figure}

Since our primary goal is the stationary current across the chain, we shall analyse the matrix $\widetilde{R}$ in the basis of the eigenstates of the current operator,
\begin{equation}
\label{b4}
\widehat{I}=\sum_{j=1}^{\cal N} \sigma_j |\Phi_j\rangle\langle\Phi_j| \;.
\end{equation}
Two examples of the matrix $\widetilde{R}$ in this basis are given in Fig.~\ref{fig3} for $U=0$, left panel, and $U=J$, right panel. A qualitative difference between these two cases is clearly visible from the plot. In the next paragraph we quantify this difference. We conclude the present paragraph by displaying the commutation relation between the current operator and the BH Hamiltonian for $U=0$,
\begin{equation}
\label{b45}
-i[\widehat{H},\widehat{I}] - \frac{(\hat{n}_L-\hat{n}_1)}{2}=0 \;,
\end{equation}
which we shall use later on.


\begin{figure}[b]
\includegraphics[width=8.0cm,clip]{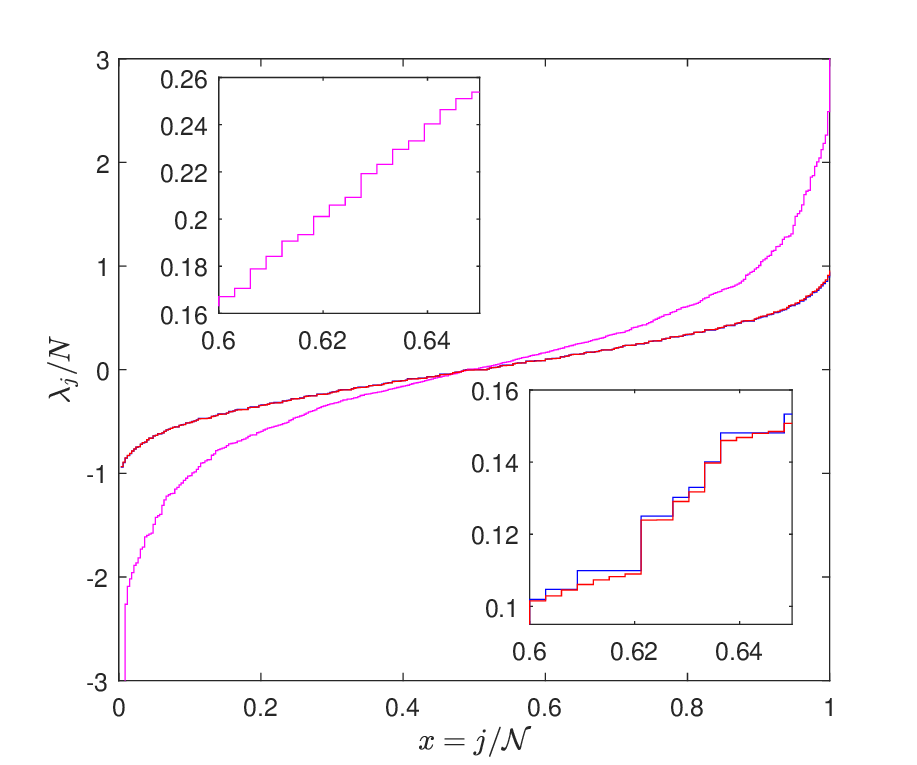}
\caption{The scaled eigenvalues of the matrix $\widetilde{R}$ for $U=0$, red line, and $U=1$, magenta line, compared to the eigenvalues of the current operator, blue line. Parameters are $L=8$, $N=4$ (${\cal N}=330$), $J=1$, and $\Gamma=0.04$.}
\label{fig1}
\end{figure} 

{\em 4.} Knowing that the BH Hamiltonian (\ref{a1}) exhibits transition to quantum chaos as $U$ is increased, we expect a similar transition for the non-equilibrium density matrix. This expectation is supported by  the visual analysis of the matrices depicted in Fig.~\ref{fig3} and results of the relevant studies on the boundary driven spin chains \cite{Pros13,Znid13}. Following Ref.~\cite{Pros13,Znid13} we consider the spectrum and eigenstates of the non-equilibrium density matrix,
\begin{equation}
\label{b6}
\widetilde{R}=\sum_{j=1}^{\cal N} \lambda_j |\Psi_j\rangle\langle\Psi_j| \;.
\end{equation}
In what follows we restrict ourselves by the case $\Gamma \ll J$. Then, if $U=0$, the states $|\Psi_j\rangle$ practically coincide with the eigenstates of the current operator $|\Phi_j\rangle$, while the eigenvalues are related to each other as  
\begin{equation}
\label{b7}
\lambda_j \approx \sigma_j / 4{\cal N} \;.
\end{equation}
To see that let us scale the density matrix $\widetilde{R} \rightarrow 4{\cal N} \widetilde{R}$ and set $\Delta\Gamma=0$ in Eq.~(\ref{b2}). Then the obtained algebraic equation differs from the commutation relation Eq.~(\ref{b45}) by a small term $\sim\Gamma$ which can be taken into account perturbatively. As expected, application of the perturbation theory removes degeneracies of the eigenvalues of the current operator, see the lower inset in Fig.~\ref{fig1}.

\begin{figure}
\includegraphics[width=8.0cm,clip]{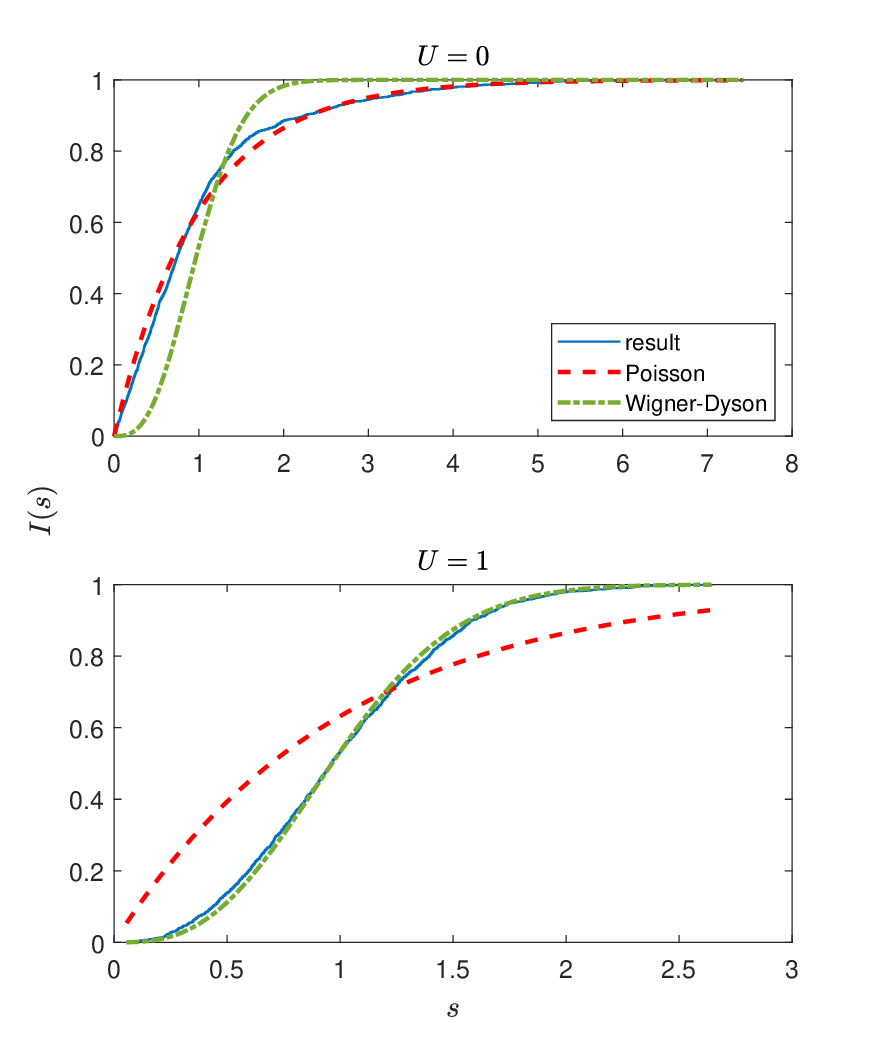}
\caption{The integrated level spacing distribution  compared to the integrated Poisson and GUE Wigner-Dyson distributions, $L=10$, $N=5$ (${\cal N}=2002$), and $\Gamma=0.04$. For the statistical analysis we took 60 percents of eigenvalues from the central part of the spectrum.}
\label{fig4}
\end{figure}

The magenta staircase curve in Fig.~\ref{fig1} depicts the case $U \ne 0$. Here we see that the width of the spectrum increases as $U$ is increased. The main difference is, however, in the spectrum statistics. Figure \ref{fig4} shows the distribution of the the scaled spacings $s=(\lambda_{j+1}-\lambda_j) f(\lambda_j)$, $f(\lambda)$ being the mean density of states, as compared to the Poisson and Wigner-Dyson distributions \cite{Wign51}. For $U=J$ an excellent agreement with the Wigner-Dyson distribution, which is the hallmark of quantum chaos \cite{Haak91,Stock99}, is noticed.


{\em 5.} Next we analyse the stationary current $I$ across the chain,
\begin{equation}
\label{b9}
I = {\rm Tr}[\widehat{I}\widehat{R}] =\sum_{j=1}^{\cal N} \lambda_j \langle \Psi_j | \widehat{I} |\Psi_j\rangle
\equiv \sum_{j=1}^{\cal N} \lambda_j I_j \;.
\end{equation}
In the case of vanishing inter-particle interactions one derives by using Eq.~(\ref{b7}) the following semi-analytic equation, 
\begin{equation}
\label{b8}
I = 4J N^2 \Delta\Gamma \int_0^1 \sigma^2(x) dx \;,
\end{equation}
where $x=j/{\cal N}$ and $\sigma(x)$ is the inverse function to the integrated density of states of the current operator, which interpolates the blue and red lines in Fig.~\ref{fig1}. Thus, as it is intuitively expected, for $U=0$ the total current {\em increases} with the number of particles in the system. 

\begin{figure}
\includegraphics[width=9.0cm,clip]{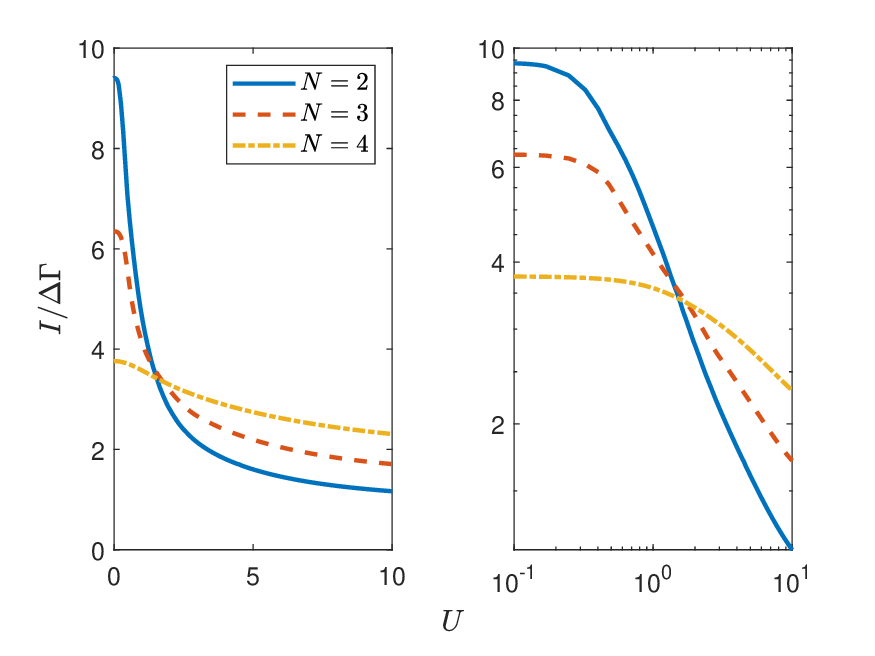}
\caption{The total current for $U=0$ as function of the interaction constant $U$ for three different values of the particle number $N=2,3,4$ in the linear and logarithmic scales, $L=10$ and $\Gamma=0.04$}
\label{fig5}
\end{figure}

The case $U\ne0$ is more subtle. Figure \ref{fig5} shows numerically obtained dependence of the stationary current on the interaction constant $U$ for $L=6$ and different number of particles $N$. One clearly identifies in this figure the critical $U_{cr}=U_{cr}(\bar{n} )$, $\bar{n}=N/L$, above which the current  drastically decreases. This critical interaction marks the crossover from the Poisson to Wigner-Dyson spectrum statistics for the non-equilubrium density matrix $\widetilde{R}$. An unexpected result is that for $U\gg U_{cr}$ the current {\em decreases} with the number of particles. Furthermore, we find that for these large $U$ the spectrum statistics is again Poissonian. 

\begin{figure}
\includegraphics[width=8.0cm,clip]{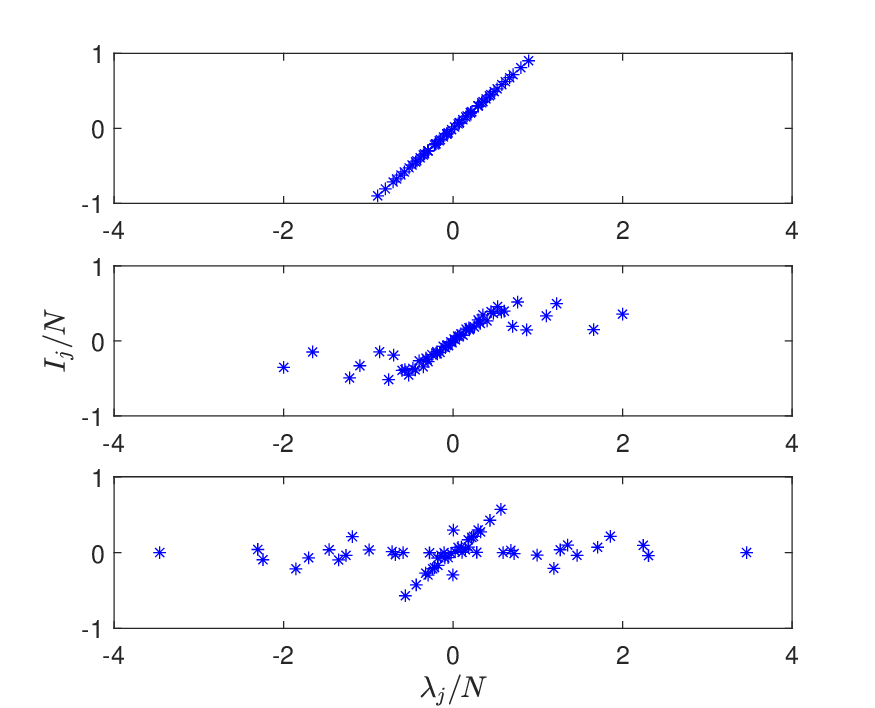}
\caption{The quantities $I_j=\langle \Psi_j | \widehat{I} |\Psi_j\rangle$ for $U=0$ (a) 0.5 (b), and 10 (c). The system parameters are $L=6$, $N=3$, and $\Gamma=0.04$.}
\label{fig6}
\end{figure}

{\em 6.} We relate the observed change of the spectrum statistics and the counter-intuitive dependence of the current on the particle number to the interaction-induced localization of the eigenstates $|\Psi_j\rangle$ and the fermionization of the strongly interacting Bose particles \cite{Pare04}. Indeed, the obvious consequence of the eigenstate localization is that the mean $I_j=\langle \Psi_j | \widehat{I} |\Psi_j\rangle$ tends to zero and is strictly zero if all bosons occupies a single site of the chain. Figure \ref{fig6} shows the quantiles $I_j$ for $(L,N)=(6,3)$ and $U=0,0.5,10$. It is seen that the fraction of the delocalized states which support the current decreases in favor of the localised states for which $I_j\approx0$. For example, in Fig.~\ref{fig6}~(c) the states corresponding to the minimal and maximal eigenvalues are the Fock states $|0,0,0,0,0,3\rangle$ and $|3,0,0,0,0,0\rangle$, respectively. Along with the localised and partially localised states one can see in Fig.~\ref{fig6}~(c) a number of the delocalized states. A closer inspection of these states shows that they are a superposition of the Fock states where occupation numbers of the chain sites are either zero or unity. Since this subspace of the Hilbert space is the Hilbert space of the hard-core bosons, we conclude that the residual conductivity of the system at large $U$ is mainly due to the hard-core bosons. As known, the spectral and transport properties of the hard-core bosons are similar to those of the non-interacting fermions and, hence, they can support ballistic transport for arbitrary large $U$ if $N/L<1$. 
We also mention that the appearance of the localised states and the integrable `hard-core boson states' is consistent with the observed change of the level-spacing distribution from the Wigner-Dyson distribution back to the Poisson distribution in the limit of large $U$. 


{\em 7.} In summary, we introduced the model for quantum transport of Bose particles across the Bose-Hubbard chain which conserves the number of particles in the chain. Similar to the standard transport model where Bose-Hubbard chain connects two particle reservoirs with different chemical potentials and where the number of particles is not conserved, the introduced model shows different transport regimes depending on the ratio between the tunnelling and interaction constants in the Bose-Hubbard Hamiltonian (\ref{a1}). Namely, for $U\sim J$ the stationary current of the Bose particles is drastically suppressed as compared to the case $U=0$. In our previous publication \cite{116} we explain this effect by transition to chaotic dynamics of the classical counterpart of the system.
In this work we use the genuine quantum approach where the object of interest is the non-equilibrium many-body density matrix of the bosonic carriers in the chain. Up to the best of our knowledge this is the first example where the non-equilibrium density matrix is calculated/analysed for the non-integrable bosonic system. (For transport properties of the fermionic and spin systems we refer the reader to the recent review \cite{Bert21}.) We found the spectrum of this matrix exhibits a transition from a regular spectrum for $U \ll J$, which obeys the Poisson statistics, to an irregular spectrum for $U\sim J$, which obeys the Wigner-Dyson statistics. In this sense we confirm the conjecture of Ref.~\cite{116} that the drastic reduction of the current for $U\sim J$ is due to the transition to quantum chaos.

Within the framework of the introduced model we also observed a new effect, -- the residual conductivity due to fermionization of the Bose particles. We notice that this is a pure quantum effect which cannot be addressed by using the classical (mean-field) or pseudoclassical (truncated Wigner function) approaches. 

{\em Acknowledgment.} 
This work has been supported by Russian Science Foundation through Grant No. N19-12-00167.

\bibliography{mybib}



\end{document}